\documentclass{svjour3}                     
\smartqed  
\usepackage{graphicx}
\def\be{\begin{eqnarray} &&}
\def\nonu{\nonumber \\ &&}
\def\ee{\end{eqnarray}}
\def\psla{\rlap \slash}
\def\bew{\begin{widetext}}
\def\ew{\end{widetext}}

\newcommand{\mbf}[1]{\mathbf{#1}}

\def\Dslash{\raise.15ex\hbox{/}\kern-.7em D}
\def\Pslash{\raise.15ex\hbox{/}\kern-.7em P}

\usepackage{color}

\renewcommand{\thefootnote}{\fnsymbol{footnote}}
\renewcommand{\bar}[1]{\overline{#1}}

\begin{document}

\title{
Projecting  the Bethe-Salpeter Equation onto the Light-Front and back: A Short Review
\thanks{Intl.
Workshop on Relativistic Description of Two- and Three-body
 Systems in Nuclear Physics
        ECT* Trento, 19 - 23 October 2009}
\thanks{CNPq and FAPESP grants are acknowledged.}
}

\titlerunning{Projecting BS equation onto LF}

\author{Tobias Frederico
\and Giovanni Salm\`e 
}

\authorrunning{Tobias Frederico and Giovanni Salm\`e }

\institute{T. Frederico \at
              Instituto Tecnol\'ogico de Aeron\'autica, DCTA, 12.228-900, S\~ao Jos\'e dos Campos, SP, Brazil \\
              Tel.: +55-12-39475939\\
              Fax: +55-12-39476960\\
              \email{tobias@ita.br}           
           \and
           G. Salm\`e \at
Istituto Nazionale di Fisica Nucleare,
Sezione di Roma, P.le A. Moro 2,
I-00185}

\date{Received: date / Accepted: date}

\maketitle

\begin{abstract}
The technique of projecting the four-dimensional two-body
Bethe-Salpeter equation onto the three-dimensional Light-Front
hypersurface, combined with the  quasi-potential approach, is briefly
illustrated, by placing a particular
 emphasis on
the relation between  the
projection method and the effective dynamics of the valence
component of the Light-Front wave function.
Some details on how to construct
 the Fock expansion of both i) the Light-Front effective interaction
and ii) the electromagnetic current operator, satisfying the proper
Ward-Takahashi
identity, will be presented, addressing the relevance of the
Fock content in the operators living onto the
Light-Front hypersurface.
 Finally, the generalization of the formalism to the
three-particle case will be  outlined.
\end{abstract}

\section{Introduction}
\label{intro}
In his 1949 seminal paper \cite{dirac}, P.A.M. Dirac proposed  three
 peculiar representations of the
Poincar\'e group, in strict relation with the choice of possible
space-time hypersurfaces without a time-like direction, (see also,
e.g., \cite{stern}): the Instant form, the Light-Front (LF) form and
the Point form (cf \cite{point} for a recent review). Each
hypersurface leads to a specific, but equivalent, description of the
dynamics of a relativistic interacting system. As a matter of fact,
the choice of an initial hypersurface, in which the points are
separated by space-like distances and therefore no causal
connections are allowed,  suggests the set of dynamically
independent variables, suited for implementing a description of the
initial state of an interacting system. In conclusion, Dirac
indicated  the potentially fruitful role of the Hamiltonian approach
for relativistically describing interacting systems, within a field
theoretical framework, as well (see, e.g., \cite{stern}).

In the Instant form, corresponding to the choice of a constant-time
hypersurface in the Minkowski space ($x_0=0$), translations and
rotations, being the generators in the stability set of the chosen
hypersurface, commute with the Hamiltonian. The eigenstates of the
Hamiltonian, constructed in terms of the Fock-space basis (in this
basis  the free Hamiltonian is diagonal), have
 both the  eigenvalues of the
 total three-momentum and the total angular moment as good quantum numbers.
 Translations and rotations
  do not mix different Fock-components of the wave
function and therefore a truncation in the Fock-space, that is
required for practical applications, is stable under those
transformations. Differently, the  Instant-form boosts have
 dynamical nature and therefore they mix
the Fock components. The shortcoming of a truncated Instant-form
Fock basis is made manifest by the lack of Lorentz invariance, e.g.
when we evaluate the expectation value of
  observables involving
initial and final rest-frames, having a relative non zero velocity.

In the LF form, where the hypersurface $x^+=t+z=0$   is chosen for
quantizing the theory, one has a kinematical subgroup of the Poincar\'e
group, built by seven generators  (see e.g.
\cite{stern,kp,karmaprep,brodsky}). Such  operators, keeping invariant the LF
hypersurface,
do not contain the
interaction and they correspond to i) three LF translations, ii) a longitudinal LF
boost along
the z-direction, iii) a rotation around z, and iv) two transverse
LF boosts, suitable linear combinations of transverse Instant-form  boosts
and rotations.
 The remaining three generators
 have a dynamical nature: i) two LF transverse rotations, like the LF boosts
 but with different sign combinations, and  ii)
the Hamiltonian $P^-=P^0+P^3$, i.e., the   generator
of LF-time $x^+$ translations. From this classification,
 immediately an
important feature stems out: the stability under LF boosts of any truncated
Fock expansion of the physical states (eigenstates of
the squared mass Fock-space operator, see, e.g., \cite{brodsky}), since the
LF boosts are diagonal in the  Fock space \cite{prlwilson}.
This property is fundamental for a consistent treatment of the boosts
when a truncated Fock basis is adopted. For
example, if one calculates electromagnetic form factors
for momentum transfers along the  $z-$direction ( $q^+>0$
and $\vec q_\perp=0$ \cite{LPS98}), the initial and final
rest-frames are related simply by a kinematical transformation, and therefore
one could use a truncated description of the initial and final states.
 As a last remark, it is worth noting that the change to LF variables allows
one to linearize the dependence upon the dynamical variable (conjugated to the
LF time), with a
simplification in the analysis of the poles in the propagators (i.e., $k^2_0-(m^2+
|{\bf k}|^2)+i
\epsilon \to k^+~[k^-  -(m^2+|{\bf k}_\perp|^2)/k^+] +i\epsilon$).

The physical outcome of a quantum theory does not depend on the
particular space-time hypersurface adopted for quantizing, and therefore
the theory has to have an equivalent four-dimensional (4D)
formulation, where explicit covariance is built in. In this respect,
the Bethe-Salpeter equation (BSE) is a 4D
field-theoretical tool \cite{grossbook} to explore the
non perturbative physics of bound and scattering states of
few-particle systems. Within a LF framework, it can be  easily shown that the
projection onto the LF hyperplane of the BS
amplitude is proportional to the {\em valence component} of the physical state
(see,
e.g., \cite{karmaprep}).  This suggests to take the BSE as the
starting point for an  investigation
of  the internal dynamics, alternative to the one where
  a coupled-equation system, generated by the Hamiltonian acting on the full Fock
space, is considered. In order to pave the way from  the
BSE to  the eigenequation for the 3D valence component 
  and back, another important
suggestion is offered by 
 the idea of ''iterated resolvents'' of Refs.
\cite{pauli1,pauli2}, where it is proposed that the complexity of
the Fock-space LF Hamiltonian  builds an effective dynamics
for the 3D valence state. Indeed, it should be pointed out that
the mass operator
for the 3D valence component can be   exactly obtained from the BSE by a
quasi-potential (QP) expansion \cite{wolja}, as investigated in Refs.
\cite{sales00,sales01} and \cite{hierareq} (where the associated set of coupled
equations for LF Green's functions was derived). 

 The QP approach makes
feasible
to single out the ''trivial'' global propagation of the interacting system by
means of a well-defined auxiliary Green's function (see Eqs. (\ref{2.11a}) and
(\ref{g0tilde}), below). This fundamental goal, in turn,
allows one to reconstruct the BS amplitude from the 3D valence component, with a
one-to-one correspondence, and  moreover, to obtain the expansion of  
the 3D effective interaction, entering
the eigenequation for the valence component. Each contribution to the expansion
immediately acquaints a transparent physical interpretation (given the strong
analogy between the LF evolution and the non relativistic one) that produces a
straightforward guidance in evaluating the relative importance of the associated
diagrams. The same analysis seems more involved in the  studies of the
Hamiltonian in a truncated Fock-space, i.e. in the analysis  of the generated 
 system of 
coupled equations. 
Noteworthy, the
convergence of the expansion, that has a direct impact on the lost of covariance
with respect to the subset of the dynamical transformations, has been
investigated in a simple bosonic Yukawa model in Ref. \cite{sales00}, where the
Fock content of each contribution has been recognized as the ordering 
''parameter''. In closing, the QP technique appears so appealing within the LF
framework, that it could be very interesting to implement comparisons between 
actual
calculations performed in other relativistic QP approaches, like the Covariant
Spectator Theory (see e.g. \cite{Gross2010}) applied in many relevant few-nucleon
problems.

The LF projection of the  BSE for
few-particle systems, and  the consequent truncation of the
 physical state in the Fock space,  can be useful if
there is  a dominant valence
state, or if  the normalization can be saturated at large extent by including
only few Fock components. In Nuclear Physics, the nucleonic component is
largely dominant, as in the deuteron case, while in applications to
 hadrons, Fock components beyond the valence one have been
recognized to be relevant, in particular in the description  of
physical quantities pertaining to inelastic channels, like
  structure functions and generalized parton distributions (from deeply
virtual Compton scattering \cite{diehl}).  Recently, a signature of the
 the relevance
of the components beyond the valence one  has been singled out in
the evaluation of the nucleon electromagnetic form factors by using
constituent quark degrees of freedom in the $\vec q_\perp=0$ frame
\cite{LPS98}. As a matter of fact,  a zero in  the proton electric
form factor has been associated with a cancellation between valence
and nonvalence contributions to the electromagnetic current (see
e.g. \cite{nucleonplb09} and \cite{pucket} for the experimental
overview).

Aim of the present review is to yield  some insights on i) the
technique for  projecting the 4D BSE onto the three-dimensional (3D)
LF hypersurface using the quasi-potential approach and ii) the
construction of the effective dynamics  of the valence component of
the LF wave function for both two-  \cite{sales00,sales01} and
three-particle systems  \cite{adneipos}. Moreover, some details on
how to obtain
 an effective electromagnetic current operator i) acting on the valence
 component of
composite two-boson and two-fermion systems,  and ii) fulfilling the
Ward-Takahashi identity (WTI)
\cite{adnei07} and \cite{adnei08}, are given.

The review is organized as follows.
In Sec. 2 the QP reduction is presented as a tool to
eliminate the relative LF-time. In Sec. 3, the relation one-to-one
between the BS amplitude
and the LF
valence wave-function is illustrated. In Sec. 4, a hierarchy of
coupled equations for LF Green's
functions is discussed in order to show
 the Fock content of the projection
of the BS equation onto the LF hyperplane. In Sec. 5
the projection technique for  three-body BS
equations  is briefly discussed, along with  the role of
induced three-body forces. In Sect. 6, the
electromagnetic current operator, acting on the valence component
and fulfilling the  LF Ward-Takahashi identity is introduced, with some remarks
 on  both structure and covariance of the
truncated (in the Fock space) current.
In Sec. 7,  a summary and  some perspectives are presented.

\section{The Quasi-Potential reduction: a tool for eliminating the relative LF-time}
\label{QPred} The 4D BSE (see, e.g. \cite{grossbook}) is the starting
point of many studies of few-body systems which aim to account for
relativity. The BSE requires a relativistic field-theoretical
approach, based on an interacting Lagrangian  able to model the
system under investigation. In the particular case of a two-body
scattering, the 4D scattering equation for the transition matrix,
$T(K)$, with total four-momentum $K$ is written as follows
\begin{eqnarray}
T(K) =  V(K)+V(K){\cal G}_0(K)T(K) ~, \label{1}
\end{eqnarray}
where the interaction $V(K)$ contains, in principle, all the possible two-body
irreducible diagrams. The two-body disconnected Green's function,
${\cal G}_0(K)$, should include self-energy terms, but they are neglected
 in the
approach developed so far. Therefore, in the case of two bosons, ${\cal G}_0(K)$ becomes
\begin{equation}
G_0(K)=\frac{\imath^2}{2\pi}\frac 1{\widehat{k}_1^2-m_1^2+i\varepsilon }\frac
1{\widehat{k}_2^2-m_2^2+i\varepsilon }, \label{2a}
\end{equation}
where
$\widehat{k}_i^\mu $ is the four-momentum operator and
the factor $2\pi$ is introduced for convenience.
For two fermions one has ${\cal G}_0(K) \to  G^F_0(K)
 = \left(\widehat{{\rlap\slash k}}_1 +m_1\right)
\left(\widehat{{\rlap\slash k}}_2+m_2\right)
G_0(K).$

The two-particle bound-state with total 4-momentum $K_B$,
$K_B^2=M_B^2$, corresponds to a T-matrix pole. The residue is
associated with the vertex function,  namely the nontrivial of part the
BS amplitude, $\Psi_B$,
solution of the following homogeneous equation
\begin{eqnarray}
\left| \Psi_B  \rangle \right.= {\cal G}_0(K_B)V(K_B)\left| \Psi_B \rangle
\right. \ . \label{1.2ab}
\end{eqnarray}
 The normalization
condition~\cite{grossbook} has to be satisfied in order to fully determine  $\left|
\Psi_B \rangle \right.$. For  scattering states with total
4-momentum $K$, the BS amplitude is a solution of the following inhomogeneous
equation
\begin{eqnarray} \left| \Psi^+
\rangle \right.= \left|\Psi_0\rangle\right.+{\cal G}_0(K)V(K)\left| \Psi^+
\rangle \right.\ , \label{1.2as}
\end{eqnarray}
while the T-matrix, solution of Eq.(\ref{1}), corresponds to the
connected four-point function which brings information on both the
scattering and bound states. In Eqs. (\ref{1.2ab}) and (\ref{1.2as})
the  four-momentum
conserving $\delta$-function is factorized out.

Since in the LF projection method a central role is played by the on-minus-shell propagation, as it
will be emphasized below, let us write the relevant matrix elements of the free two-boson
Green's function, viz
\be
\left\langle k_1^{\prime -}\right| G_0(K)\left|k^-_1\right\rangle
= \nonu =\frac{\imath^2}{2\pi}
\frac{\delta \left(k^{\prime -}_1-k^{-}_1\right)}{\widehat k%
^+_1 (K^+-\widehat k^+_1) \left(k_1^--\widehat k_{1on}^-+\frac{i\varepsilon}{\widehat k^+_1}\right)
\left(K^--k_1^--\widehat k_{2on}^-+\frac{ i\varepsilon}{K^+-\widehat k^+_1} \right)}~,
\label{3}
\ee
where  the LF  four-momenta are  $k_i=(k^-_i:=k^0_i-k^3_i\ ,
\ k^+_i:=k^0_i+k^3_i \ , \ \vec k_{i\perp})$, $\widehat
k^-_{ion}=(\widehat{\vec k}_{i\perp}^2+m^2)/ \widehat k^+_i$
($i=1,2$) is the on-minus-shell
momentum operator, with eigenfunctions given by  the LF plane waves,   $ \langle x^-_i \vec
x_{i\perp}\left| k^+_i\vec k_{i\perp}\right\rangle= {\cal N}~
e^{-\imath(\frac12 k^+_ix^-_i-\vec k_{i\perp} . \vec x_{i\perp})}$.
 The  completeness relation and the normalization are
\begin{eqnarray} \int \frac{dk^+d^2k_\perp}{2(2\pi)^3} \left. |k^+\vec
k_\perp\rangle \langle k^+\vec k _\perp \right. |
=\mbox{\boldmath$1$} , \label{nboson}\end{eqnarray} and $\langle
k^{\prime +}\vec k^\prime_\perp |k^+\vec k_\perp
\rangle=2(2\pi)^3\delta(k^{\prime +}-k^+)\delta(\vec
k^{\prime}_\perp-\vec k_\perp)  $, respectively.

The free two-fermion propagator, $G^F_0(K)$, can be decomposed in an on-minus-shell term,
${\overline
G}_0(K)$, and a part
that contains the so-called instantaneous (in the LF-time) contribution, since the
 Dirac propagator can be separated in two terms as follows
\begin{eqnarray}\frac{\psla{k}+m}{k^2-m^2+i\varepsilon}=
\frac{\psla{k}_{on}+m}{k^+(k^--k^-_{on}+ {i\varepsilon\over
k^+})}+\frac{\gamma^+}{2k^+} \ . \label{instant}
\end{eqnarray}
where the first term  yields the on-minus-shell propagation, while the second one
   the LF-time instantaneous term
of the Dirac propagator. For the fermion case, we will be interested in
the  analogous of Eq. (\ref{3}),  written in terms
of  ${\overline
G}_0(K)$, namely
\begin{eqnarray}
\left\langle k_1^{\prime -}\right| \overline
G_0(K)\left|k^-_1\right\rangle =  \left( \widehat{\rlap\slash
k}_{1on} +m_1\right) \left(\widehat{\rlap\slash k}_{2on} +
m_2\right) \left\langle k_1^{\prime -}\right|
G_0(K)\left|k^-_1\right\rangle \ ,
\label{3b}\end{eqnarray}
 For the sake of a
unified formal treatment of two-boson and two-fermion systems, {\it
in what follows  we put } $\overline G_0(K)\equiv  G_0(K)$.

One could easily extend the present analysis to
systems composed by
particle-antiparticle   or by other mixtures, like  a fermion and a boson
(see e.g. \cite{mathiotvertex}).

The first step for projecting the BSE onto the LF surface is the introduction
of
the free-resolvent, i.e., the Fourier transform in $K^-$ of the
global $x^+$-time free propagator of the two-particle system. This  amounts
to
  integrate the matrix elements, Eq. (\ref{3}) (or Eq. (\ref{3b})), of the 4D $G_0(K)$
over $k^-_1$ and $k^{\prime-}_1$, so that   the relative LF
time between the particles is eliminated, and one remains with a dependence upon $K^-$, i.e.
\begin{eqnarray}
| G_0(K)|:=  \int dk^{\prime -}_1 dk^{ -}_1
\left\langle
k_1^{\prime -}\right|  G_0(K)\left|k^-_1\right\rangle\equiv g_0(K)
\label{2.11a}
\end{eqnarray}
where $g_0(K)$, called the free-global LF propagator, is a 3D operator depending upon the LF
momenta $(k^+_i,\vec k_{i\perp })$ only,   and it is explicitly given by
\begin{eqnarray}
g_0(K)&=& \frac{%
\widehat P}{\widehat{k}_{1}^{+}(K^{+}-\widehat{k}_{1}^{+})\left( K^{-}-\widehat{%
k}_{1on}^{-}-\widehat{k}_{2on}^{-}+i\varepsilon\right) } \ ,
\label{2.11}
\end{eqnarray}
where $\widehat P=i\theta (K^{+}-\widehat{k}_{1}^{+})\theta (\widehat{k}_{1}^{+})$ for two-bosons and
$$\widehat P= i\theta (K^{+}-\widehat{k}_{1}^{+})
\theta (\widehat{k}_{1}^{+})(2m_{1})(2m_{2})\Lambda _{+}(\widehat{k}_{1on})\Lambda _{+}
(\widehat{k}_{2on})$$ for two-fermions. The projector
$\Lambda_{+}(\widehat{k}_{on}) =\left( \widehat{\rlap\slash
k}_{on}+m\right) /2m$ is the positive energy spinor projector. A positive value for
 $K^+$  is used without any loss of generality.

In Eq. (\ref{2.11a}), the vertical bars ''$|$ ''on the right side and  on
the left one indicate that the minus components  in $|k^-\rangle$
and $\langle k^{\prime -}|$ have to be integrated
out \cite{sales00,sales01}, respectively. Notice that, for  two
fermions,  the inverse of $g_0(K)$ exists only in the valence
sector, since the projectors, $\Lambda_+$, single out only positive
energy states.

Within the  QP approach \cite{wolja}, where  an auxiliary
interaction, $W(K)$, is introduced, the full T-matrix  is solution
of the following coupled equations \begin{eqnarray} && T(K)
=W(K)+W(K)\widetilde{G}_{0}(K)T(K),  \label{2.1}
\\
&& W(K)=V(K)+V(K)\Delta_0(K)W(K)\ , \label{2.3a} \end{eqnarray}
where $ \Delta_0(K)=G_{0}(K)-\widetilde{G}_{0}(K) $  for bosons and
$\Delta_0(K)=G^F_{0}(K)-\widetilde{G}_{0}(K) $ for fermions. The
auxiliary Green's function
 $\widetilde G_{0}(K)$
is a 4D operator, depending upon the four-momenta of the two
constituents, and  it represents the key quantity of the LF
projection.  It is the 4D image of the 3D dimensional $g_0(K)$,
that, we  strongly stress, does not contain the relative-time
propagation of the system. Therefore $\Delta_0(K)$ just takes into
account such a propagation in the 4D space, namely it will be an
essential ingredient in the description of
 the internal dynamics of the
systems.
The 4D operator $\widetilde G_0(K)$ is  defined by
\be
\widetilde G_0(K)=\bar\Pi _0(K) g_0(K) \Pi_0(K)   ~, \label{g0tilde}
\ee
where \be  \bar\Pi _0(K)=  G_0(K) |~g^{-1}_0(K)~~,  \quad \quad
\Pi_0(K)=g^{-1}_0(K) ~|  G_0(K). \ee These operators,
{\em the free reverse LF-projection operators}, connect three and four dimensional quantities. In
the next section the corresponding interacting operators will be
introduced.  It is worth noting that the choice of $\widetilde G_0(K)$ and the corresponding
integral equation for $W(K)$, Eq. (\ref{2.3a}), allows only for
LF two-body irreducible terms.

The solution by iteration of Eq. (\ref{2.3a}) is given by \be W(K)=\sum_{n=1}^\infty W_n(K),
\label{expw}\ee
with $W(K)_n=V(K)~(\Delta_0(K)~V(K))^{n-1}$. The diagrammatic analysis of
the series (\ref{expw}) shows  that for each term one has a specific
Fock content, associated to  the propagation of  the virtual intermediate
 states, as discussed to some extent in Sect. \ref{hiera}. Of course,
 a truncation of the sum in Eq. (\ref{expw}) puts a bound on
  the number of the
Fock components involved in the actual calculation. Moreover, the same holds for
the 3D
effective interaction
$w(K):={\Pi}_{0}(K)W(K)\overline\Pi_{0}(K)$ that determines the valence
component, as shown in the following Section. In particular,
the 3D effective interaction, since it is non diagonal in the Fock space, makes possible the coupling of the valence
sector to the higher Fock components of the wave function (formal details
are  given in Sec. 4 and \cite{hierareq}).

It is worth noting that, in model studies (see
\cite{sales00,miller1,miller2,miller3,carbonellepja1,carbonellepja2}),
the expansion   (\ref{expw}) is
rapidly converging, since
the probability of higher Fock states is  quickly decreasing
\cite{karmanovnpb1,karmanovnpb2}.

 In fermionic models, with
point-like couplings, the LF BSE was investigated by retaining the
lowest order kernel and subtle problems, related to the divergences
produced by the dependence upon  the transverse momentum, were found
\cite{glazek,carbonell1,carbonell2}. Part of the difficulties can be
ascribed to the absence of  the instantaneous terms as analyzed in
Refs. \cite{sales01,bakker07}.

Following  Refs. \cite{sales00,sales01,adnei07,adnei08}, one can construct a 3D LF T-matrix, $t(K)$,
from the
4D one. In particular, one has
\begin{equation}
t(K)={\Pi}_{0}(K)T(K)\overline\Pi_{0}(K) = w(K)+w(K)g_{0}(K)t(K)=w(K)+w(K)g(K)w(K)\label{3.1}
\end{equation}
where $g(K)$ is the interacting LF Green's function, fulfilling the integral
equation
\be
g(K)=g_0(K)+g_0(K)w(K)g(K), \label{LFRESOLV0}
\ee
Notice that it also holds $g(k)=g_0(k)+g_0(k)t(K)g_0(k)$.
The 3D operator, $g(K)$, is the Fourier transform in $K^-$ of
the global LF-time propagator, viz
\be  g(K)= | G_0(K)|+ |
G_0(K)T(K) G_0(K)|, \label{greenproj}
\ee
and evolves the system from an initial state, defined on a given LF
hypersurface, to another one, after a LF-time interval $x_f^+-x_i^+>0$.
By  iterating once the integral equation (\ref{2.1}), and
using Eqs.(\ref{g0tilde}) and (\ref{3.1}), one has
\be
T(K)=W(K)+W(K)\left[ \widetilde{G}_{0}(K)+\widetilde{G}_{0}(K)T(K)
\widetilde{G}_{0}(K)\right] W(K)=
\nonu
=W(K)+W(K) \bar\Pi_0(K) g(K)\Pi_0(K) W(K)\ , \label{3.2}
\ee
This relation allows one to map the 3D dynamics into the 4D space.

It turns out that
the on-mass-shell matrix elements of $T(K)$, which define the
two-constituent scattering amplitude, are identical to the ones obtained
from the on-minus-energy-shell matrix elements of $t(K)$ (see
discussion in \cite{sales00,sales01}). Unless otherwise indicated,
 the operators $g_0(K)$ and $w(K)$ have to be evaluated
with a ''$+\imath \varepsilon$'' prescription.

\section{ The BS amplitude and the LF valence component}
\label{lfval}
The relation between the BS amplitude $\left|\Psi_B\right\rangle$
and the valence component of the LF wave function, $\left| \Phi
_{B}\right\rangle$, for a bound state with total momentum $K_B$ is
given by~\cite{sales01} (let us recall that we formally put $G_0(K)\equiv \overline G_0(K)$)
\be
\left| \Psi _{B}\right\rangle ={\cal G}_{0}(K_{B})W(K_{B}){G}
_{0}(K_{B})|~g_0^{-1}(K_B)\left| \Phi _{B}\right\rangle \ ,
\label{3.6a}
\ee
where the valence component is the
solution of the following eigenequation
\be
\left| \Phi _{B}\right\rangle =g_{0} (K_{B})w(K_{B})\left| \Phi
_{B}\right\rangle \ , \label{lfbse}
\ee
It should be pointed out that, in the case of fermions, the instantaneous terms from the Dirac
propagators appear in i) ${\cal G}_0\equiv G_0^F$, ii) $W$ and iii) the
effective interaction $w$.

The identity $ G_0(K_B)|~\left(g^{-1}_{0}
(K_{B})-w(K_B)\right)\left| \Phi _{B}\right\rangle =0$ can be added
to Eq. (\ref{3.6a}) in order to get
\be
\left| \Psi _{B}\right\rangle
=\left[1+\Delta_0(K_B)W(K_{B})\right]
{G}
_{0}(K_{B})|~g_0^{-1}(K_B)\left| \Phi _{B}\right\rangle \ ,
\label{psicov}
\ee
This expression holds not only for bound states but also for scattering states (see
e.g. \cite{adnei08}), viz
\be
\left| \Psi ^+\right\rangle=\Pi(K)~\left| \Phi^+\right\rangle \
\label{psicov1}\ee
where $\Pi(K)$, {\em the interacting reverse LF-projection operator}, is given by
\be
\Pi(K):=\left[1+\Delta_0(K)W(K)\right]{G}
_{0}(K)|~g_0^{-1}(K)=G_R(K)|~g^{-1}(K)~~.\label{invop}
\ee
with
\begin{eqnarray}
G_R(K):= G_0(K)+ {\cal G}(K)V(K) G_0(K)
={\cal G}(K){\cal G}_0^{-1}(K)G_0(K)    ~ ,\label{GR}
\end{eqnarray}
Notice that in $G_R(K)$  the on-minus-shell-Green's function
$G_0(K)$ appears on the
rightmost position, and this leads to apply the  "$|$" operation on the right.

The operator $\Pi(K)$ acts on the valence component of the
LF wave function in Eqs.
(\ref{psicov}) and (\ref{psicov1}) and it allows one
to fully reconstruct the 4D BS
amplitude for both bound and scattering states, starting from the valence
 wave
function. The LF-conjugated operator, $\bar\Pi(K)$ is given by
\be
\bar \Pi(K):=g^{-1}(K)~|G_L(K)
\ee
with
$
G_L(K):= G_0(K)+ G_0(K)V(K){\cal G}(K)= G_0(K){\cal G}_0^{-1}(K){\cal G}(K)  ~ ,\label{GL}
$
 that allows the  "$|$"
operation on the left. The reverse LF projectors make compact
 the relations between operators and states living onto a 3D
hypersurface and the full 4D counterparts.  For instance the relation between the BS amplitude and the
valence component can be written as
$
|\Psi \rangle
=\Pi(K)|\Phi\rangle
$ and
 $\left\langle\Psi\right|=
 \left\langle\Phi\right |\bar \Pi(K)
$.
These relations can be applied  to both two-boson \cite{sales00,sales01} and two-fermion
systems~\cite{adnei07,adnei08} with the proper choice of ${\cal G}$, ${\cal G}_0$ and the
on-minus-shell $G_0(K)$. Reversely, the valence component of the
LF wave function can be obtained directly from the BS
amplitude by using Eqs. (\ref{invop}) and  (\ref{GR}) \cite{sales00,sales01,adnei07,adnei08}
\be
 | G_0(K) {\cal G}_0^{-1}(K) \left|\Psi\right\rangle =
~| G_0(K) {\cal G}_0^{-1}(K){\cal G}(K){\cal G}^{-1}_0(K) G_0(K)
|~g^{-1}(K)\left|\Phi\right\rangle = \nonu=\left|\Phi\right\rangle \
. \label{psiproj} \ee To conclude this section, it is worth noting
that the 3D valence LF wave functions
 are  solutions of the squared mass eigenvalue
equation:
\begin{eqnarray}
g^{-1}(K )\left| \Phi \right\rangle =0~, \label{valeneq}
\end{eqnarray}
with suitable boundary conditions for bound and scattering
states, respectively. In particular, the LF scattering state is the
solution of the inhomogeneous equation,
\begin{eqnarray}
\left|\Phi^{+}\right\rangle=\left|\Phi_0\right\rangle+g_0(K)w(K)
\left|\Phi^{+}\right\rangle,\label{scattlf}
\end{eqnarray}
with outgoing boundary condition.

\section{Hierarchy equations for LF Green's functions}
\label{hiera}
In order to gain deep insights in
 the Fock content of the dynamical quantities involved in the approach
 presented in the previous
Sections, one can combine  the investigation of the Fock structure of a
Hamiltonian (actually a hadronic Hamiltonian)
performed in Refs. \cite{pauli1,pauli2}, with
 the LF projection technique \cite{hierareq}. For that purpose, the free
resolvent  is rewritten as
\begin{eqnarray}
g_0(K)=| G_0(K)|:=  \int dk^{\prime -}_1 dk^{ -}_1
\left\langle k_1^{\prime -}\right|
G_0(K)\left|k^-_1\right\rangle \
=i\widehat\Omega^{-1}g_0^{(2)}(K)\widehat\Omega^{-1} \ , \label{eq5}
\end{eqnarray}
where the phase space operator  is conveniently defined by
$\widehat\Omega:=\sqrt{\widehat k_1^+(K^+-\widehat k^+_1)}$.
It should be pointed out that $\widehat\Omega$ makes the
LF projection ''$|$''
invariant with respect to the kinematical subgroup of the Poincar\'e group.
For
spinless particles, the free two-body LF Green's function
is a particular case of the N-body LF Green's function given by
\begin{eqnarray}
g^{(N)}_0(K)=  \left[\prod _{j=1}^N \theta (\widehat k_j^{+})\theta
(K^{+}-\widehat k_j^{+})\right]\left( K^{-}-\widehat K_0^{(N)-}+i\varepsilon
\right)^{-1} ,
\label{gomega}\end{eqnarray}
 where $\widehat K_0^{(N)-}=\sum_{j=1}^N\widehat{k}_{jon}^-$ is
the free LF Hamiltonian.

Let us consider a two-boson system. By introducing the operator
$\widehat\Omega$, the interacting LF Green's function, Eq.
(\ref{LFRESOLV0}), can be rewritten as follows \be g^{(2)}(K)=
g^{(2)}_0(K)+g^{(2)}_0(K) \nu (K)   g^{(2)}(K) \  , \ee where
$g^{(2)}\equiv -i\widehat\Omega g(K)\widehat\Omega$, $ \nu (K)=i
\widehat\Omega^{-1} w(K)\widehat\Omega^{-1}$. From Eq. (\ref{expw}),
the leading and next-to-leading order contributions to $\nu(K)$  are
given by \be \nu^{(2)}(K)=i \left[\widehat\Omega g_0(K)\right]^{-1}|
G_0(K)V(K) G_0(K)| \left[g_0(K)\widehat\Omega\right]^{-1}\ ,
\label{eq14}
\\ \nonumber &&\\  &&
\nu^{(4)}(K) =i \left[\widehat\Omega g_0(K)\right]^{-1}| G_0(K)V(K)
G_0(K)V(K) G_0(K)|\left[g_0(K)\widehat\Omega\right]^{-1} \nonu
-i\left[\widehat\Omega g_0(K)\right]^{-1}| G_0(K)V(K)\widetilde
G_0(K)V(K) G_0(K)|\left[g_0(K)\widehat\Omega\right]^{-1} \ .
\label{eq16} \ee Notice that $\nu^{(4)}(K)$ is two-body irreducible,
due to the subtraction of the last term in Eq. (\ref{eq16}). The
content of the operator $\nu^{(n)}$ in the LF Fock-space can be
investigated, within a ladder approximation, in a Yukawa bosonic
Lagrangian model, ${\cal L}^B_I=g_S\phi _1^{\dagger }\phi _1\sigma
+g_S\phi _2^{\dagger }\phi _2\sigma$ with $\phi_1$, $\phi_2$ and
$\sigma$ bosonic fields. Then, the
  interaction vertex operator, acting between Fock states
differing by one quantum $\sigma$, has matrix element given by e.g.,
\be \langle q k_\sigma |v|k\rangle= -2(2\pi)^3\delta^3 (\tilde
q+\tilde k_\sigma-\tilde k) \frac{g_S}{\sqrt{q^+k^+_\sigma k^+}}
\theta (k^+_\sigma)~ , \label{eq18} \ee where the LF momenta are
indicated by the convention: $\tilde q\equiv \{q^+,\mbf
{q}_\perp\}$. The states are normalized according to (\ref{nboson}).
The effective interaction $\nu (K)$, up to next-to-leading order in
$v$ can be obtained from the suitable LF-time ordered diagrams,
since the LF projections, ''$|$'', allows one to play a  game
analogous to the case of the
 non relativistic
perturbation theory in the Fock space. Then, one gets \cite{pauli1,pauli2,hierareq}
\begin{eqnarray}
\nu (K)\approx \nu ^{(2)}(K)+\nu^{(4)}(K)=vg_0^{(3)}(K)v
+vg_0^{(3)}(K)vg_0^{(4)}(K)vg_0^{(3)}(K)v \ . \label{eq21}
\end{eqnarray}
 By reminding that in the Yukawa model, the BSE kernel $V(K)$ contains two
interaction vertexes, $v$, according to Eqs. (\ref{eq14}) and
Eq. (\ref{eq16}) one can perform the following identifications: i) $\nu^{(2)}\equiv
vg_0^{(3)}(K)v$  and ii)
$\nu^{(4)}\equiv vg_0^{(3)}(K)vg_0^{(4)}(K)vg_0^{(3)}(K)v$. Such expressions
straightforwardly show the Fock content of each term. In particular, the presence
of $g_0^{(3)}$ and $g_0^{(4)}$ points to an intermediate propagation of three
and four bosons, respectively.
 Once the previous analysis is performed at any order, it turns
out that  in general $$\nu(K)= vg^{(3)}(K)v\ . $$ In turn, from Eq.
(\ref{eq21}), one sees that $g^{(3)}$ is coupled to the four-body
Green's function, which should be coupled to the five-body one, and
so on. By an obvious generalization, one can construct a hierarchy
of coupled equations for the interacting bosonic Green's functions,
viz \be g^{(2)}(K)=g^{(2)}_0(K) +g^{(2)}_0(K)v
g^{(3)}(K)vg^{(2)}(K)~ \ . \ . \ . \nonu
 g^{(N)}(K)=g^{(N)}_0(K) +g^{(N)}_0(K)v g^{(N+1)}(K)vg^{(N)}(K)~
 \ . \ . \ . \ \label{eq22}
\ee
Those coupled equations encode the full Fock-space content of the QP
expansion, within  the LF projection framework.

An analogous study can be carried out for the two-fermion system, by adopting
the Yukawa model given by ${\cal
L}_{I}^F=g_{S}\overline{\Psi_1 }\Psi_1\sigma+g_{S}\overline{\Psi_2
}\Psi_2\sigma$, with $\Psi_1$ and $\Psi_2$ being the fermionic fields.
The interaction vertex operator, acting between Fock states differing by
zero, one and two $\sigma$'s,  has matrix elements given by
\be
\langle (q,s^\prime) k_\sigma |v|(k,s)\rangle= -2m(2\pi)^3\delta^3
(\tilde q+\tilde k_\sigma-\tilde k) \frac{g_S}{\sqrt{q^+k^+_\sigma k^+}} \theta
(k^+_\sigma){\overline u}(q,s^\prime)u(k,s)~  \label{eq18a} \\ &&
 \langle (q,s^\prime)k^\prime_\sigma|v|(k,s)k_\sigma\rangle=
-2(2\pi)^3\delta ^3(\tilde q+\tilde k^\prime_\sigma-\tilde k-\tilde k_\sigma)\delta_{s^\prime s}
\frac{g^2_S}{\sqrt{k^{\prime +}_\sigma k^+_\sigma}} {\theta
(k^{\prime +}_\sigma)\theta (k^+_\sigma) \over
 k^++k_\sigma^+} ~\label{eq18b} \\
&&\langle (q,s^\prime) k^\prime_\sigma k_\sigma|v|(k,s) \rangle=
-2(2\pi)^3\delta^3 (\tilde q+\tilde k^\prime_\sigma+\tilde k_\sigma-\tilde k )\delta_{s^\prime s}
\frac{g^2_S}{\sqrt{k^{\prime +}_\sigma k^+_\sigma}} {\theta
(k^{\prime +}_\sigma)\theta (k^+_\sigma) \over
 k^+-k_\sigma^+} ~ \label{eq18c}
\ee
for fermion states properly normalized. The
instantaneous terms in the two-fermion propagator give origin to
Eqs. (\ref{eq18b}) and (\ref{eq18c}). Since  $\nu(K)$ has
terms that couple sectors of the Fock space that differ at most by
two sigma's \cite{sales01}, then one gets the following expression
for the coupled set of Green's
functions
\begin{eqnarray}
&&g^{(2)}(K)=g^{(2)}_0(K)+g^{(2)}_0(K)v\left[
g^{(3)}(K)+g^{(4)}(K) +g^{(3)}(K)vg^{(4)}(K) \right.\nonumber  \\
&&+\left. g^{(4)}(K)vg^{(3)}(K)\right]vg^{(2)}(K)~,  \ . \ . \ .
\nonumber \\
&&g^{(N)}(K)=g^{(N)}_0(K) +g^{(N)}_0(K)v \left[
g^{(N+1)}(K)+g^{(N+2)}(K) +
g^{(N+1)}(K)vg^{(N+2)}(K) \right. \nonumber  \\
&&+\left.g^{(N+2)}(K)vg^{(N+1)}(K)\right]vg^{(N)}(K)~,  \ . \ . \ .
\label{hfer}
\end{eqnarray}

It is important to notice that
truncating in the Fock space  the effective interaction $\nu$
is different
from
truncating  the coupled set of Eqs. (\ref{eq22}) or
(\ref{hfer}). This can be easily understood by considering
the two-boson case and restricting
the intermediate-state propagation up to four-particles, namely retaining up
to $g^{(4)}_0(K)$. Then,  one gets
 $g^{(2)}(K)\simeq g^{(2)}_0(K) +g^{(2)}_0(K)v g^{(3)}(K)vg^{(2)}(K)$
with $g^{(3)}(K)\simeq g^{(3)}_0(K) +g^{(3)}_0(K)v
g^{(4)}_0(K)vg^{(3)}(K)$, where one has propagations up to four particles,
 given
the presence of  $g^{(3)}(K)$. On the other side, from Eq. (\ref{eq21}),
one has
$vg_0^{(3)}(K)v$, without higher Fock propagations.

\section{LF projection of three-body BS equations}

The approach briefly revised in Sects. \ref{QPred} and \ref{lfval}
has been extended to three-particle
systems in Ref. \cite{adneipos}. In the last years, within a LF framework,
3-body systems have been  i) investigated within zero-range models
\cite{fred92,karmanovzr}, ii) applied to the description of the
nucleon \cite{wilson} and iii) adopted for analyzing the
 quark mass effects in heavy
baryons  \cite{suisso}. It is very important to notice that
the QP expansion allows one to systematically deal
 with higher Fock-state contributions  to the
dynamics of the three-body valence component, in fully analogy with
the two-body case.

 The starting point of the investigation
 performed  in Ref. \cite{adneipos}
is
the three-body Bethe-Salpeter equation for the transition-matrix, within the
ladder approximation, viz
\begin{equation}
T=V+V{\cal G}_0T~~~;~~~V=\sum
V_i~~~~;~~~V_i=V^{(2)}_{jk}S^{-1}_i,\label{BSE}
\end{equation}
where $V^{(2)}_{jk}$ stands for an interaction between particles $j$
and $k$ corresponding to 2-body irreducible diagrams. $S_i$ is the
individual particle propagator and ${\cal G}_0=S_1S_2S_3$. {\it The complexities
produced  by the spin degrees of freedom are omitted in what follows}.

The same QP formalism shown in Eqs. (\ref{2.1}) and (\ref{2.3a})
 can be applied to the three-body BSE, but with the obvious extension to the
 three-body case of the free Green's function, namely
 ${\cal G}_0=S_1S_2 \to {\cal G}_0=S_1S_2S_3$. In order to get the three-body LF Green's
 function $g_0(K)$, one has to
   project onto the LF hyperplane
by integrating over two independent minus momentum
components, i.e.
$\tilde{G}_0:=G_0||g_0^{-1}||G_0~$ with $g_0:=||G_0||$ and $G_0$ the generalization to
the three-particle case of the on-minus-shell two-body propagators (see Eq. (\ref{3})
for bosons
and Eq. (\ref{3b}) for fermions).
The double bar ''$||$ ''operation on the right or on the left means
\be ||G_0:=\int dk_1^-dk_2^-\langle k_1^-,k_2^-|~G_0,~ ~G_0||:=\int
dp_1^-dp_2^- ~G_0~|p_1^-,p_2^-\rangle.
\ee

The Faddeev decomposition of
the QP as $W_i= V_i+ V_i\Delta_0 \sum_j W_j $,
 leads to the components of the effective interaction
$w_i:=  \Pi_0W_i\overline\Pi_0:=g_0^{-1}||G_0W_iG_0||g_0^{-1}.$ The free reverse
LF time operators for the three-body system,
$\Pi_0$ and $\overline\Pi_0$, are introduced in analogy to the
two-body case, with the difference that they now contain a double
integration. Finally
three-body LF transition matrix reads as follows
\begin{eqnarray}
t:=\Pi_0T\overline \Pi_0=\sum_{i=1}^3w_i+\sum_{i=1}^3w_ig_0t~~,
\end{eqnarray}
where the standard Faddeev decomposition, $t=\sum_{i=1}^3 t_i$,
leads to a couple set of equations
$
t_i=\overline t_i +\overline t_i g_0(t_j+t_k)
$
with $ \overline t_i= (1-w_ig_0)^{-1}w_i$.

One is tempted to identify $ \overline t_i$ with the two-body
subsystem 3D transition matrix, however this is not
possible. It contains irreducible three-body terms from the point of
view of the LF global propagation. To clarify this point, let us
consider an example with the  interaction Lagrangian density:
$\mathcal{L}_I= \sum_{i=1}^3 g \phi^\dagger_i\phi_i\sigma$, where
three different spin zero bosons exchange a scalar quantum $\sigma$
Moreover, let us simplify our discussion assuming that $V_i$ corresponds
to the 4D ladder one-boson exchange. The expansion of
$W_i$, e.g., up to next-to-leading order is given by
\begin{eqnarray}
w_i= \Pi_0V_i\overline \Pi_0+\Pi_0V_i\Delta_0V_i\overline \Pi_0+
\Pi_0V_i\Delta_0(V_j+V_k)\overline \Pi_0+\cdots \ ,
\end{eqnarray}
The leading-order $\Pi_0V_i\overline \Pi_0$ corresponds to a LF
 two-particle interaction in the presence of the $i$-th boson,
and it  is built through the coupling of  three- and
four-particle Fock sectors. Pictorially, one has an intermediate
propagation of a   a four-particle state,
i.e. three $\phi_i$ bosons and an exchanged quantum $\sigma$, between an
initial and final
three-boson free propagations.
 The term $\Pi_0V_i\Delta_0V_i\overline \Pi_0$ corresponds to
two-body stretched boxes involving two boson out three, with the third one, the
$i$-th boson, acts
as a spectator.
The term  $\Pi_0V_i\Delta_0V_j\overline \Pi_0$ corresponds to an {\em induced
three-body force} due to the elimination of the relative LF-time
between the particles. It should be pointed out that
the induced three-body forces, from the 4D point of view,
 are quite different
from the intrinsic
three-body forces, and play a very important role in the determination of the
three-body dynamics,
as shown   by Karmanov and Maris  in \cite{karmanovmaris}, where
the calculation of the three-boson bound-state mass has been presented.

\section{Electromagnetic Current and LF Ward-Takahashi Identity}
The electromagnetic current operator plays a central role for the phenomenology,
and therefore it deserves a detailed analysis within any relativistic framework.
In particular,  one should pay attention  to the fulfillment of the
Ward-Takahashi
identity (WTI)~\cite{gross}, as well. For a system of two charged particles 1
and 2, in the Minkowski space, the WTI reads as follows
\begin{eqnarray}
Q_\mu{\mathcal J}^\mu(Q)=\left[{\cal G}^{-1},\widehat e_1\right]
+(1\leftrightarrow 2), \label{jwti}
\end{eqnarray}
where the charge operator for particle $i$ has matrix elements given by
$\langle k_i|\widehat
e_i|p_i\rangle=e_i\delta^4\left(k_i-p_i-Q\right).$
The full Green's function of the interacting two-particle system is
a solution of $ {\cal G}(K)={\cal G}_0(K)+{\cal G}_0(K)V(K){\cal G}(K)$. The current operator may contain
two terms, a free contribution and an interacting part, viz \be {\mathcal
J}^\mu(Q)={\mathcal J}^\mu_0(Q)+{\mathcal J}_I^\mu(Q) \ , \ee
It is worth noting that the free term, ${\mathcal J}^\mu_0(Q)$,
leads to the impulse approximation
 derived by
Mandelstam \cite{mandelstam}, where self-energy insertions/vertex corrections
were disregarded.

Once the relation between the BS amplitude and the
3D LF valence component (see Sect. \ref{lfval})
has been established, the LF em current operator can be constructed from
the matrix element of the 4D current (see
\cite{adnei07,adnei08}). For both scattering and bound states
one has
\begin{eqnarray}
\left \langle\Psi_{f}\right|{\mathcal J}^\mu(Q)\left|
\Psi_{i}\right\rangle=\langle\phi_{f} |j^\mu(K_f,K_i) |\phi_i\rangle
~,\label{lfc1}
\end{eqnarray}
where  $ j^\mu(K_f,K_i)$ is the 3D LF current, that  acts on the valence wave
functions and
{\it includes two-body operators}. It is  given by
\begin{eqnarray}
j^\mu(K_f,K_i) :=  \overline \Pi(K_f){\mathcal J}^\mu(Q)\Pi(K_i)
~.\label{lfc2}
\end{eqnarray}
A 3D LF electromagnetic
current operator for two-boson interacting systems, acting on the valence state
and fulfilling a WTI,  has  been also obtained
by using the method of gauging equations \cite{kvi03}. The
 achieved result is  in agreement with
 Eq. (\ref{lfc2}).

In order to  implement the WTI for the 3D current  one starts with
the corresponding relation for the 4D current, Eq. (\ref{jwti}), and
then one applies the reverse LF projector, as suggested by Eq.
(\ref{lfc2}). The WTI satisfied by $j^\mu(K_f,K_i)$  is given by \be
Q_\mu j^\mu(K_f,K_i) = \overline \Pi(K_f)\left[{\cal
G}^{-1}(K_f)\widehat e-\widehat e{\cal G}^{-1}(K_i)\right]\Pi(K_i)
~,\label{lfc3.1} \ee with $\widehat e=\widehat e_1+\widehat e_2$.
After applying some formal manipulations, illustrated in  great
detail in \cite{adnei08}, one gets
 \be Q_\mu j^\mu(K_f,K_i) =  g^{-1}(K_f)~\widehat {\cal Q}^L_{LF}- \widehat {\cal
Q}^R_{LF}~ g^{-1}(K_i) \label{lfc11}  \ee
where there are the fully interacting 3D LF Green's functions, labeled by the total
  initial and final
momenta,  and  the {\it left} and {\it right} LF charge operators
have been introduced, according to the following definition \be
\widehat {\cal Q}^L_{LF}=|{ G}_0(K_f){\cal ~G}^{-1}_0(K_f)~\widehat
e~ \Pi(K_i)= |{ G}_0(K_f)~{\cal G}^{-1}_0(K_f)~\widehat e~
\Pi_0(K_i) \ , \label{eilf} \ee \be \widehat {\cal
Q}^R_{LF}~=\overline \Pi(K_f)~\widehat e ~{\cal G}^{-1}_0(K_i)~{
G}_0(K_i)| =\overline \Pi_0(K_f)~\widehat e ~{\cal G}^{-1}_0(K_i)~{
G}_0(K_i)|
 \label{eflf} \ .\ee
{ with the same formal assumption indicated below Eq. (\ref{3b}).}
 In the case of fermions,
the explicit expression, e.g. for particle 1, is given by
\begin{eqnarray} &&\widehat {\mathcal Q}^L_{1LF}= \Lambda_+(\widehat
k_{1on})\frac{m_1}{\widehat k^{+}_1}\gamma_1^+~\widehat e_{1LF}
\Lambda_+(\widehat k_{1on})\Lambda_+(\widehat k_{2on})
~,\label{lfc9}  \\
&&\widehat {\mathcal Q}^R_{1LF}= \Lambda_+(\widehat k_{1on})~\widehat
e_{1LF}\frac{m_1}{\widehat k^{+}_1}\gamma_1^+  \Lambda_+(\widehat
k_{1on})\Lambda_+(\widehat k_{2on}) ~,\label{lfc10}
\end{eqnarray}
where the notation $\widehat e_{1LF}$ indicates
the 3D LF counterpart of the 4D operator $\hat e$,
with matrix elements given by
\begin{eqnarray}
\langle k^{\prime+}_1,{\vec k}^\prime_{1\perp}|\widehat
e_{1LF}|k_1^+,\vec k_{1j\perp}\rangle:=e_1\delta^3\left(\tilde
k^{\prime}_1-\tilde k_1-\tilde Q\right)~ .
 \label{d3}
\end{eqnarray}
The normalization condition can be obtained sandwiching
 the operator $\gamma^+ m/k^+$  between LF spinors.

In the case of  a spin-zero boson system, one simply has
$\widehat {\mathcal Q}^R_{1LF}\equiv \widehat {\mathcal Q}^L_{1LF}= \widehat
e_{1LF}$.

It is very important to stress that the LF charge operator are interaction free,
within the approach of Refs. \cite{adnei07,adnei08}, where no
self-energy corrections were included.

From Eq. (\ref{lfc11}), current conservation straightforwardly follows by
taking the matrix elements between   3D LF valence components,
 solutions of the wave equation (\ref{valeneq}) and
noting that the {\it left} and {\it right} charge operators do not
contain any $\imath\varepsilon$ dependence.

By multiplying both the left and right hand sides of Eq.~(\ref{lfc11}) by
$g(K_f)$ and $g(K_i)$, respectively, one gets
\begin{eqnarray}
Q_\mu g(K_f)j^\mu(K_f,K_i)g(K_i) =  \widehat{\mathcal Q}^L_{LF}g(K_i)-
g(K_f)\widehat {\mathcal Q}^R_{LF}~,\label{lfc12}
\end{eqnarray}
which corresponds to the LF projection of the 5-point function without
instantaneous terms in the external legs in the case of fermions.
For bosons this identity was also derived in \cite{kvi03}.

Within a field theoretical approach, the Poincar\'e covariance of
${\mathcal J}_\mu$ is ensured. Therefore, the covariance properties of the
 matrix elements of the
operator $j_\mu$,
are fulfilled,  since all the matrix elements of the lhs of
Eq. (\ref{lfc1}) are properly related through the Lorentz
transformations.  But for a truncated QP expansion, see  Eq. (\ref{expw}),
it is expected that  the full covariance of the
description will be  lost, at some extent.  A quantitative analysis of
such a Poincar\'e covariance breaking can be
systematically carried out by considering more and more terms in the
QP expansion. In any case, even after introducing a proper truncation, the corresponding  LF
current fulfills a WTI, where   truncated LF Green's functions appear
\cite{adnei07,adnei08}. The corresponding current conservation can be retrieved
by using the valence wave functions obtained from eigenequations with the
inverse of the truncated LF Green's function.
Although the
explicit expression of the truncated LF current is rather involved,
the physical picture that arises is
quite sensible: the truncated LF two-body current, that fulfills the proper WTI,
 is  generated by attaching the
photon in all the
possible ways to the truncated effective interaction operators, that
 are present in the
truncated reverse LF projectors (see Eq. (\ref{lfc2})) and    are
irreducible with respect to LF two-body propagation.

Within a framework kinematically Poincar\'e invariant, the rich phenomenology
of the electromagnetic processes can be addressed by adopting the approach
presented in this Section. It is interesting just to remind few issues.
In the fermionic case, the instantaneous contributions play an essential  role
at the
formal level, and therefore one should be eager to investigate possible
signatures of those terms. Furthermore, for both bosonic and fermionic systems,
the choice of a frame different from the celebrated Drell-Yan one, namely
a frame  where the plus
component of the momentum transfer is not vanishing, allows one to study the Fock
content of the system state, by coupling small components to large components
(for an analysis of the physical impact on actual cases in Hadronic Physics
see, e.g., Refs. \cite{nucleonplb09}, \cite{pacheco02} and  \cite{pasquini}).

\section{Summary and perspectives}

In this review, we have given a short presentation of the LF projection method, based on
a combination of the integration on the minus component of the constituent four-momenta and the
quasi-potential formalism \cite{wolja}. This approach
  allows one to formally establish a one-to-one relation between the Bethe-Salpeter amplitude and the
3D LF valence component of a bosonic or fermionic  system, and it makes natural
 to address the issue of the Fock content of the dynamics governing the system under investigation.
 In particular,
 in Sect. \ref{hiera}, it has been shown that by
using the quasi-potential approach, in the  spirit of the
''iterated resolvents'' suggested by H.C. Pauli \cite{pauli1,pauli2},
one can construct a set of coupled equations for the LF resolvents,
that allows one to arrange a sort of tomography in the  Fock space of
the Bethe-Salpeter equation. Finally, the extension of the LF projection method to the
 three-particle case has been outlined.

From the phenomenological point of view, the last Section  contained the most relevant topic.
There, we briefly presented the derivation of the conserved 3D LF electromagnetic current
operator and the associated Ward-Takahashi identity. The issue of the truncation in the Fock space of
the LF current, and the consequent elaboration for obtaining a truncated Ward-Takahashi identity
has been discussed. In particular, it has been pointed out that in order to fulfill the current
conservation, the valence wave functions
 have to be solutions of the suitable  eigenequation, obtained from the
truncated Green's function.

In conclusion, we would mention that, within the framework of the LF projection method,
 the calculation of the deuteron electromagnetic form factors,
 with two-body currents obtained from
 a pion exchange interaction,  and
the  investigation of the relation between the analytic properties of the Nakanishi
representation and the Fock decomposition of the BS amplitude (see e.g.
\cite{carbonellepja1,carbonellepja2}) are in progress.

\end{document}